\documentclass[amsmath,amssymb,floatfix,twocolumn,showpacs,prl,aps,superscriptaddress]{revtex4}

\usepackage{mathtools}
\usepackage[all]{xy}
\input xy
\xyoption{matrix}
\xyoption{2cell}
\xyoption{arrow}
\xyoption{curve}
\xyoption{all}

\usepackage{graphicx}
\usepackage{epsfig}
\usepackage{dcolumn}% Align table columns on decimal point
\usepackage{bm}% bold math

\usepackage{amsmath}
\usepackage{graphicx}
\usepackage[latin1]{inputenc}
\usepackage{ulem}
\usepackage{epstopdf}
\usepackage{subfigure}
\usepackage{color}

\newcommand{\abs}[1]{\left| #1 \right|}
\newcommand{\EFB}{E_{\text{\begin{tiny}FB\end{tiny}}}}

\newcommand*\xbar[1]{%
  \,
  \hbox{%
    \vbox{%
      \hrule height 0.7pt % The actual bar
      \kern0.3ex%         % Distance between bar and symbol
      \hbox{%
        \kern-0.2em%      % Shortening on the left side
        \ensuremath{#1}%
        \kern0.0em%      % Shortening on the right side
      }%
      
    }%
  }%
}

\begin{document}

%% REVISION VERSION CONTROL:
%% V1 - Daniel (all major changes, supplemental)
%% V2 - Josh (separated supplemental source tex, minor grammar, tex source cleanup)
%% SUBMISSION - Referee Report Back. 
%% V3 - Josh (first address of referee comments). 

%%%%%%%%%%%%%%%%%%%%%%%%%%%%%%%%%%%%%%%%%%%%%%%%%%%%%%%%%%%%%%%%%%%%%%%%%%%%%
\title{Flat Bands Under Correlated Perturbations}
\author{Joshua D. Bodyfelt}
\affiliation{New Zealand Institute for Advanced Study, Centre for Theoretical Chemistry \& Physics, Massey University, Auckland, New Zealand}
\author{Daniel Leykam}
\affiliation{Nonlinear Physics Centre, Research School of Physics and Engineering, The Australian National University, Canberra ACT 0200, Australia}
\author{Carlo Danieli}
\affiliation{New Zealand Institute for Advanced Study, Centre for Theoretical Chemistry \& Physics, Massey University, Auckland, New Zealand}
\author{Xiaoquan Yu}
\affiliation{New Zealand Institute for Advanced Study, Centre for Theoretical Chemistry \& Physics, Massey University, Auckland, New Zealand}
\author{Sergej Flach}
\affiliation{New Zealand Institute for Advanced Study, Centre for Theoretical Chemistry \& Physics, Massey University, Auckland, New Zealand}

\begin{abstract}
Flat band networks are characterized by coexistence of dispersive and flat bands. Flat bands (FB) are generated by compact localized eigenstates (CLS) with local network symmetries, based on destructive interference. Correlated disorder and quasiperiodic potentials hybridize CLS without additional renormalization, yet with surprising consequences: 
(i) states are expelled from the FB energy $E_{FB}$,
(ii) the localization length of eigenstates vanishes as $\xi \sim 1/\ln (E-\EFB )$, 
(iii) the density of states diverges logarithmically (particle-hole symmetry) and algebraically (no particle-hole symmetry),
(iv) mobility edge curves show algebraic singularities at $E_{FB}$.
Our analytical results are based on perturbative expansions of the CLS, and supported by numerical data in one and two lattice dimensions.
\end{abstract}


% 73.43.Cd 	Theory and modeling (Quantum Hall Effects, low dimension)
%71.23.An 	Theories and models; localized states
%71.30.+h 	Metal-insulator transitions and other electronic transitions
%71.10.-w 	Theories and models of many-electron 
%73.20.Fz 	Weak or Anderson localization
%37.10.Jk 	Atoms in optical lattices
%64.70.Rh 	Commensurate-incommensurate transitions
%05.50.+q 	Lattice theory and statistics (Ising, Potts, etc.) 

\maketitle
%%%%%%%%%%%%%%%%%%%%%%%%%%%%%%%%%%%%%%%%%%%%%%%%%%%%%%%%%%%%%%%%%%%%%%%%%%%
{\sl Introduction ---}
Disorder has a profound effect on waves in periodic potentials, smoothing out van Hove singularities in the density of states and generating Anderson localization~\cite{vanHove_occurrence_1953, anderson, kramer93}. Three dimensional disordered lattices support metal-insulator transitions and mobility edges, while in one and two dimensions the effect of disorder is much simpler, localizing all eigenstates and completely suppressing transport. 
Correlated disorder changes this picture and allows for complex behavior even in one dimension~\cite{correlated_review}. Examples include the appearance of resonant transmission channels (random dimer model~\cite{dunlap90} and tight binding models of DNA~\cite{dna1,dna2}), metal-insulator transitions (Aubry-Andr\'{e} model~\cite{AA}), and mobility edges (correlations with power law decay~\cite{lyra98,izrailev99}). Counterintuitively, certain correlations can even enhance localization~\cite{kuhl08}. Recent advances have allowed the direct observation of these fundamental effects using cold atoms~\cite{cold_atom_AL,wu2007,hyrkas13,cold_atom_review} and photonic systems~\cite{kuhl2000, dietz11,bellec13}.

The above elastic potential scattering effects can be both strongly amplified and qualitatively changed when the kinetic energy is quenched, such as in a strictly flat dispersion band~\cite{bergholtz13,parameswaran2013,richter,mielke,tasaki,bergman2008}. Flat (macroscopically degenerate) bands occur when perfect destructive interference allows for compact localized eigenstates (CLS), modes with nonzero amplitude only at a finite number of lattice sites. 
There are flexible approaches to designing flat band (FB) lattices in a variety of dimensions~\cite{goda2006,2D,richter,hyrkas13}, which can support new topological phases~\cite{bergholtz13}, and even model the fractional quantum Hall effect resulting from flat-band (FB) degeneracies of electronic Landau levels interacting within a magnetic field~\cite{parameswaran2013}. 
% Modern examples of FB lattice construction support three-dimensional (3D)~\cite{goda2006} / 2D~\cite{2D} / 1D~\cite{richter,hyrkas13} cases, the presence of new topological phases, and even model the fractional quantum Hall effect resulting from flat-band (FB) degeneracies of electronic Landau levels interacting within a magnetic field~\cite{parameswaran2013} . 

Anderson localization in flat bands displays a variety of unconventional features including inverse Anderson transitions~\cite{goda2006,Nishino}, multifractality at weak disorder~\cite{chalker10}, and effective heavy-tailed disorder distributions~\cite{leykam_flat_2013}. Recently the local symmetries of the CLS were used to detangle uncorrelated disorder into two distinct terms: one that renormalizes the energies of the CLS, and another that hybridizes them with modes belonging to other dispersive bands~\cite{flach_detangling_2014}. This detangling suggests a way to independently control the two terms using appropriately correlated potentials. Such control is feasible with ultracold atoms~\cite{cold_atom_AL,wu2007,hyrkas13,cold_atom_review} and photonic systems~\cite{kuhl2000, dietz11,bellec13}, but can be also expected for electric or sound propagation along crystal surfaces exposed to adsorbing atoms and molecules.

In this letter, we consider locally correlated disorder and quasiperiodic potentials in flat band lattices. The compact flat band states hybridize with other dispersive degrees of freedom, but their (bare) energies are not renormalized. This leads to a strong competition between the macroscopic number of compact localized states, generating new spectral singularities (in contrast to uncorrelated disorder, which smooths out all singularities). The resulting surprising action of the perturbations is that:
(i) all states are expelled from the FB energy $E_{FB}$,
(ii) the localization length of eigenstates vanishes as $\xi \sim 1/\ln (E-\EFB )$,
(iii) the density of states diverges logarithmically (particle-hole symmetry) and algebraically (no particle-hole symmetry)
for disorder potentials, 
(iv) and metal-insulator transitions induced by quasiperiodic potentials are promoted by the flat band to mobility edges, whose curves show algebraic singularities at $E_{FB}$. Thus, correlated potentials provide a way to ``fine-tune'' the flat band singularity strength, or convert it into more useful form (e.g. mobility edge). Our analytical results are based on perturbative expansions of the CLS and supported by numerical data.

\begin{figure}%[htb]
\centering
\includegraphics[width=\columnwidth]{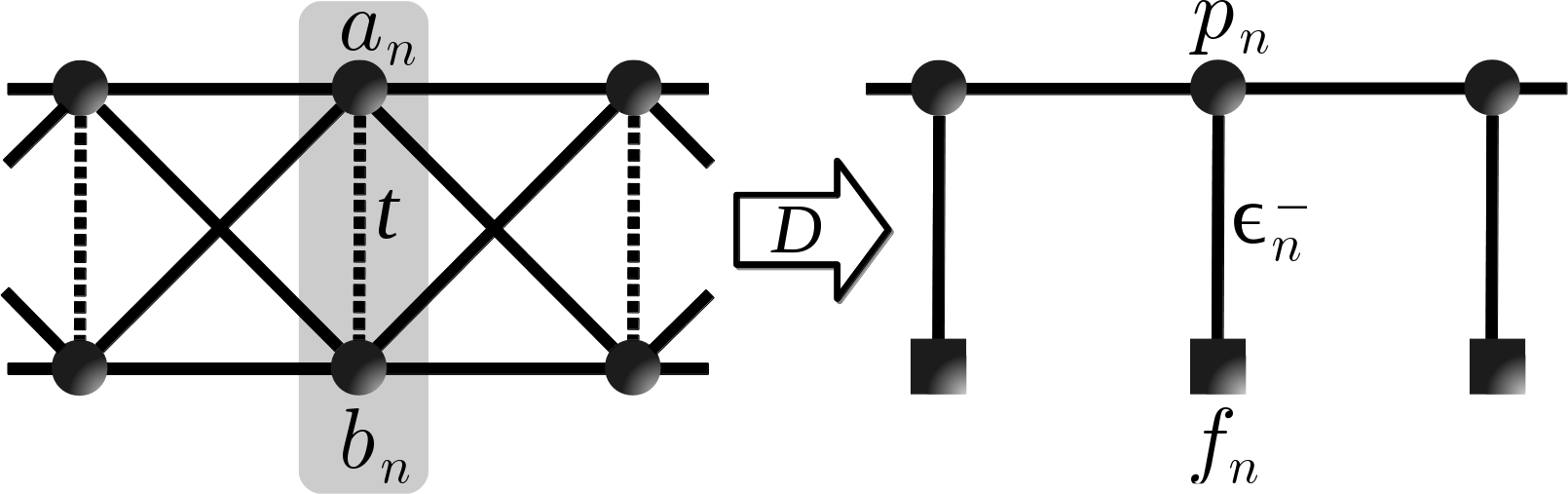}
\caption{The cross-stitch lattice structure (left) of Eq.(\ref{eq:eigen}). The detangled version of Eq.(\ref{eq:fp}) is shown in the right plot.
}
\label{fig:model}
\end{figure}

{\sl 1D Model ---} To illustrate the idea we will start with the simplest case of a one-dimensional FB model with exactly one dispersive band and one flat band. The cross-stitch lattice, shown in the left plot in Fig.\ref{fig:model}, consists of two interconnected chains. Its unit cell is given by two lattice sites shaded in 
the figure, and the wave amplitude at the cell is $\psi_n = \left(a_n, b_n\right)^T$. Stationary waves follow the eigenvalue problem
\begin{equation}
E \psi_n = \epsilon_n  \psi_n - tV \psi_n - T (\psi_{n-1} +  \psi_{n+1})\;,  \label{eq:eigen}
\end{equation}
with
\begin{equation*}
\epsilon_n = \begin{pmatrix} \epsilon_n^a & 0 \\ 0 & \epsilon_n^b \end{pmatrix} \; , \;
 V=\begin{pmatrix} 0  & 1 \\ 1 & 0 \end{pmatrix} \;, \;  T = \begin{pmatrix} 1 & 1 \\ 1 & 1 \end{pmatrix}.
\end{equation*}

In the crystalline case of $\epsilon_n = 0$, Eq.\eqref{eq:eigen} is put into a Bloch basis 
%($T_n \mapsto T_n e^{i k}$) 
and diagonalized to give the dispersion curves
\begin{equation*}
E(k) = -4 \cos(k)-t, \quad \EFB = t \;.
\end{equation*}
One band is flat and independent of $k$, with Bloch modes $B_n( k ) = (1,-1)^T e^{i k n} / \sqrt{2}$.  Due to the degeneracy, any superposition of these Bloch modes is also an eigenmode, and one can construct compact localized modes $\psi_n=(1,-1)^T \delta_{n,n_0} / \sqrt{2}$. Applying the local rotations
\begin{equation}                    
\phi_n \equiv \begin{pmatrix} p_n \\ f_n \end{pmatrix} = D \psi_n\;, \;
%\begin{pmatrix} \epsilon^{+}_n  \\ \epsilon^{-}_n\end{pmatrix} = \frac{D \epsilon_n}{\sqrt{2}} ; \quad
D=\frac{1}{\sqrt{2}} \begin{pmatrix} 1 & 1 \\  1 & -1 \end{pmatrix}.
\label{eq:rot}
\end{equation}
with  $\epsilon_n^{\pm}=(\epsilon_n^a \pm \epsilon_n^b)/2$,
Eq.\eqref{eq:eigen} becomes~\cite{flach_detangling_2014}
\begin{equation}
\begin{aligned}
(\xbar{E}+2t)\, p_n &= \epsilon_n^+ p_n + \epsilon_n^-\, f_n - 2\left( p_{n-1} + p_{n+1} \right) \\
\xbar{E}\, f_n &= \epsilon_n^+ f_n + \epsilon_n^-\, p_n, 
\end{aligned}
\label{eq:fp}
\end{equation}
where we measure the energy deviation from $E_{FB}$ as $\xbar{E}=E-t$, and the CLS $f_n$ are locally hybridized with the dispersive variables $p_n$ at strength $\epsilon_n^-$, while
their energies are renormalized exclusively through nonzero $\epsilon_n^+$ (see Fig.~\ref{fig:model} right).
Experimental realizations of the cross-stitch model can be obtained both in its original and detangled forms; the latter having a simpler geometry easily obtained using microwave resonator networks~\cite{bellec13}.
%%%%%%%%%%%%%%%%%%%%%%%%%%%%%%%%%%%%%%%%%%%%%%%%%%%%%%%%%%
%%%%%%%%%%%%%%%%%%%%%%%%%%%%%%%%%%%%%%%%%%%%%%%%%%%%%%%%%%
%%
%%
{\sl Disorder ---} 
Real systems are never perfect and experience fluctuating deviations from an ideal setup.
In Ref.~\cite{flach_detangling_2014}, a disorder potential was added assuming
onsite energies $\epsilon_n^{a,b}$ are random uncorrelated, with a probability density function (PDF) 
of finite variance
$\mathcal{P}(\epsilon)=1/W$ for $|\epsilon| \leq W/2$, and $\mathcal{P}=0$ otherwise.
Excluding the CLS variables $f_n$ from Eq.\eqref{eq:fp}, one obtains
\begin{equation}
\frac{\epsilon^p_n -\xbar{E}}{2} p_n= p_{n-1} + p_{n+1} \;,\;
\epsilon^p_n=\epsilon_n^+ + \tfrac{(\epsilon_n^-)^2}{\xbar{E} - \epsilon_n^+}-2t,
 \label{eq:tb}
\end{equation}
which is a tight-binding chain under an energy-dependent onsite disorder potential $z=\epsilon_n^p$. 
Its PDF %$\mathcal{W}_p(z) = \frac{2}{z^2} \int \mathcal{P}\left(y\right) \mathcal{P} \left( \frac{2}{z} - y \right) dy$ 
displays Cauchy tails~\cite{flach_detangling_2014} with diverging variance at the FB energy.
Consequently, at weak disorder $W \ll1$ the localization length $\xi$ of 
an eigenstate $p_n^\nu \sim e^{-\frac{n}{\xi}}$  scales as $\xi \sim 1/W^2$ away from $E_{FB}$, and as $\xi \sim 1/W$ at $E_{FB}$. This energy-dependent inverse localization length $\xi^{-1}(E)$ is numerically calculated using
the recursive iteration
\begin{equation}
\xi^{-1}(E) = \lim_{M\rightarrow+\infty}\frac{1}{M}\sum_{n=1}^M \ln \abs{\frac{p_{n+1}}{p_n}}.
\label{eq:loclength}
\end{equation}
Though the disordered FB states are much more strongly localized than other states,
their width still diverges for weak disorder. This is because the disorder is uncorrelated, so it performs both energy renormalization and hybridizaton with dispersive states at the same time.

\begin{figure}
\begin{center}
\includegraphics[width=\columnwidth]{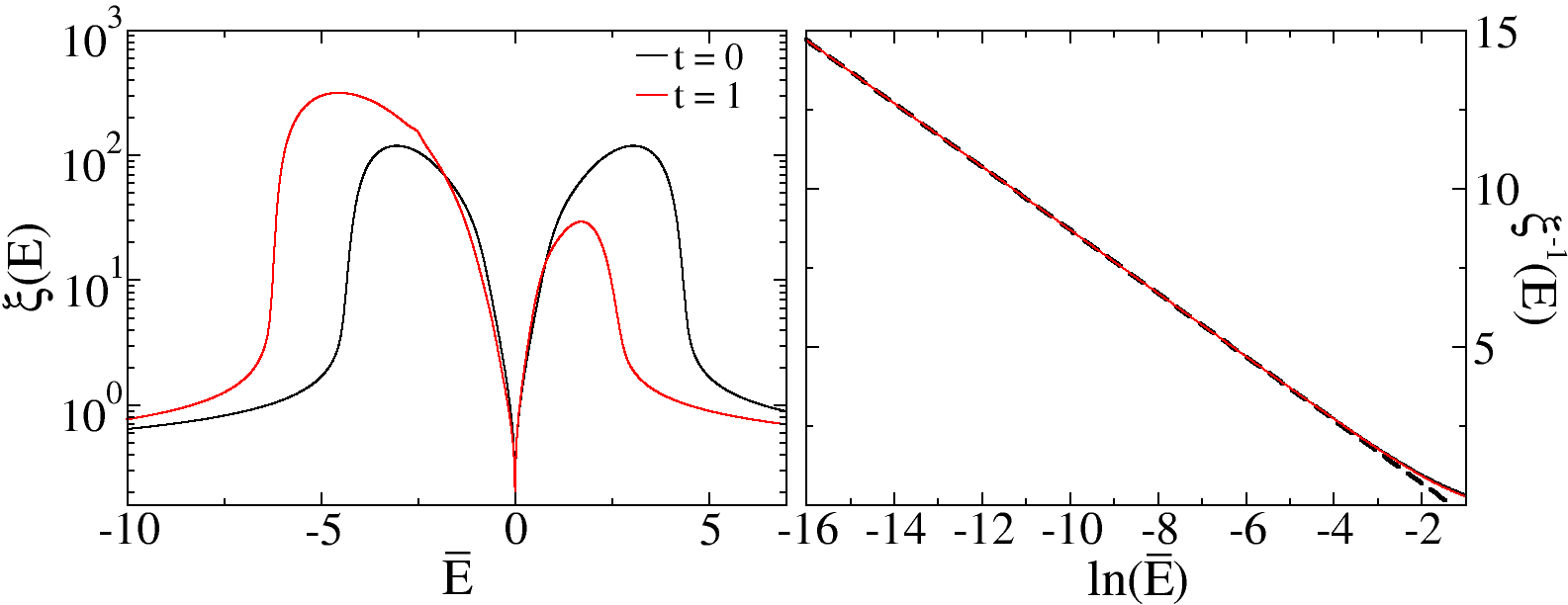}
\end{center}
\caption{Left plot: Localization length $\xi$ versus eigenstate energy $\xbar{E}=E-t$, for $t=0$ (black solid) and $t=1$ (red solid).
Right plot: Inverse localization length $\xi^{-1}$ versus $\ln \xbar{E}$ for $\xbar{E} > 0$, same color coding as in left plot.
The dashed line corresponds to Eq.\eqref{loclengthcorrdist=0}. Here, $W=4$.
}
\label{fig:LL}
\end{figure}

A drastic change occurs when the potential is correlated such that energy renormalization no longer occurs, i.e. $\epsilon_n^a=-\epsilon_n^b$, which leads to 
$\epsilon_n^+=0$ (easily implemented with microwave resonator networks~\cite{bellec13}). The remaining 
potential $\epsilon_n^-$ has PDF $\mathcal{P}(\epsilon)$, and Eq.\eqref{eq:tb} now displays a Fano 
resonance at energy $\xbar{E} = 0$ at every lattice site, which strongly scatters the dispersive degree 
of freedom $p_n$. For small $\xbar{E}$ we can neglect nonresonant terms, and 
substituting~\cite{supplemental} into Eq.\eqref{eq:loclength}, obtain the localization length
\begin{equation}
\xi^{-1} = \ln \frac{W^2}{8|\xbar{E}|} -2 .
\label{loclengthcorrdist=0}
\end{equation}
Hence irrespective of the strength $W$ of the correlated disorder, the localization length vanishes due 
to resonant scattering as the energy tends towards $E_{FB}$. We compute the localization length 
numerically using Eq.\eqref{eq:loclength}. The results in Fig.~\ref{fig:LL} (black lines) agree 
excellently with the analytical predictions (dashed line).

%%%%%%%%%%%%%%%%%%%%%%%%%%%%%%%%%%%%%%%%%%%%%%%%%%%%%%%%%%%%%%%%%%%%%%%%
%%%%%%%%%%%%%%%%%%%%%%%%%%%%%%%%%%

While this picture of a macroscopic number of Fano resonances at $\xbar{E} =0$ can intuitively explain 
the behavior of the localization length, surprisingly the flat band energy $E_{FB}$ is completely 
emptied: no eigenstate can reside there. This follows directly from Eq.\eqref{eq:fp}, which now allows 
only for a trivial solution $p_n = f_n = 0$ when $E = E_{FB} = t$. All the compact localized states have 
hybridized and shifted their energies away; however a significant fraction stay energetically close to 
$E_{FB}$, such that the density of states still diverges at $E_{FB}$. To show this, we note that close to 
resonance the eigenmodes should strongly excite the CLS, which may hybridize among themselves. The weak 
energy shifts of these states imply the existence of a small parameter, which can be used for 
perturbative calculations. We consider first $t=0$. Up to normalization,  we 
construct~\cite{supplemental} dimer-like states at energy $\xbar{E}=\pm(\epsilon_0^-\epsilon_{1}^-)/2 \ll 
W^2/4$ (position shifts can be done without loss of generality) as
$f_0=f_{1}=\pm 1$, $p_0=\pm\epsilon_{1}^-/2,p_{1} = \epsilon_0^-/2$, $p_{n \geq 2} = \pm \epsilon_0^- (2 \xbar{E})^{n-1}/(\Pi^{n}_{m=2} \epsilon_m^-)^2$, $f_{n \geq 2} = \pm 2 p_{n-1} / 
\epsilon_n^-$. 

The density of states $\rho(\xbar{E})$ for small $\xbar{E}$ follows~\cite{supplemental} from the PDF $\int_{-\infty}^{+\infty} \mathcal{P}(x) \mathcal{P}(\frac{z}{x})|x|^{-1} dx$ of  the random number $z=\epsilon_0 \epsilon_1$ as
\begin{equation}
\rho(\xbar{E}) = \frac{4}{W^2} \left( \ln \frac{W}{2} - \ln \frac{4|\xbar{E}|}{W} \right) \;. %DL: checked, OK
\label{doscorrdist=0}
\end{equation}
Despite the result that eigenstates strictly do not exist at $E_{FB}$, the density of states $\rho(\xbar{E})$ diverges logarithmically at $E_{FB}$. 

\begin{figure}%[hbt]
\begin{center}
\includegraphics[width=\columnwidth]{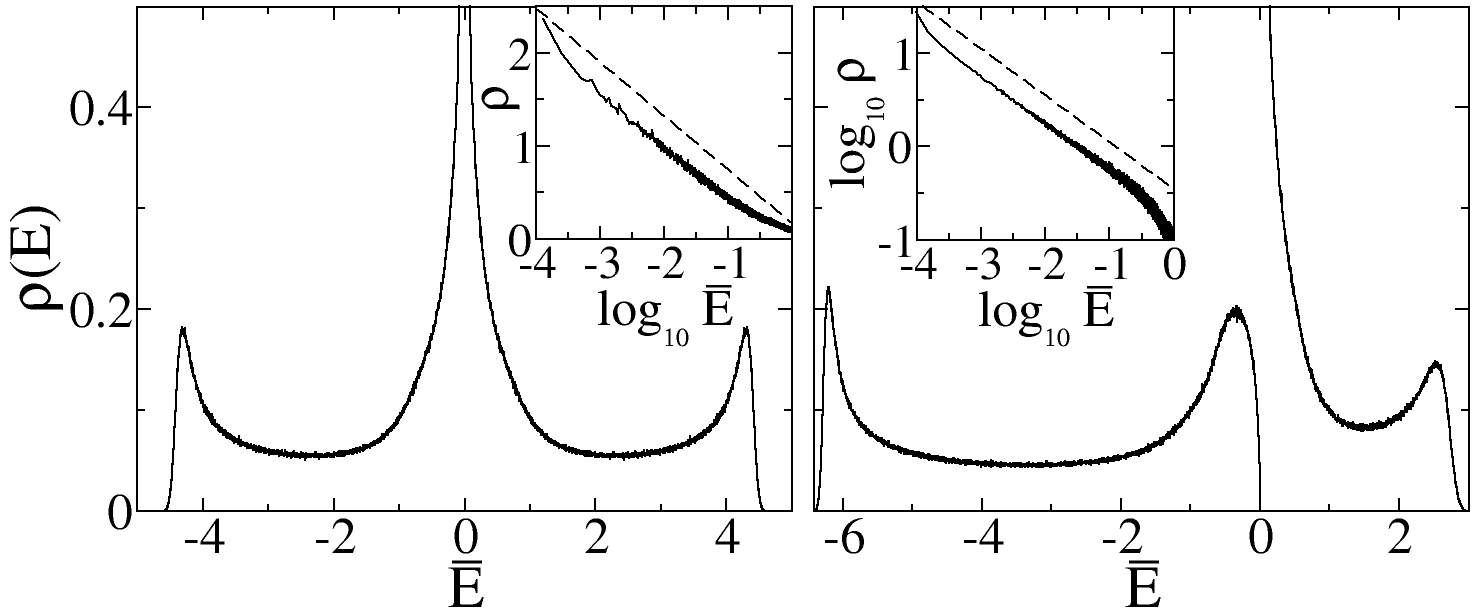}
\end{center}
\caption{
Density of states $\rho(\xbar{E})$ for $t=0$ (left) and $t=1$ (right). Divergences are observed at the flat band energies $\xbar{E} = 0$. Logarithmic scalings of positive $\xbar{E}$ in the insets describe the divergent behavior, and the dashed lines indicate theoretical results of Eqs.~(\ref{doscorrdist=0},\ref{doscorrdistneq0}). Here, $W=4$.
}
\label{fig:DoS}
\end{figure}

We perform diagonalizations of Eq.\eqref{eq:eigen} and
obtain the density of states following well-known schemes (see chapter 3 in Ref.~\cite{hjs2006}). The result in the left of Fig.~\ref{fig:DoS} confirms the predicted logarithmic divergence.
It therefore also confirms that we identified the correct group of eigenstates responsible for the divergence.

%%%%%%%%%%%%%%%%%%%%%%%%%%%%%%%%%%%%%%%%%%%%%%%%%%%%%%%%%%%%%%%%%%%%%%%%
%%%%%%%%%%%%%%%%%%%%%%%%%%%%%%%%%%

When the FB energy is shifted away from the particle-hole symmetry point $E_{FB}=t \neq 0$, the nature
of the localized states changes. At the energy $\xbar{E}=\epsilon_0^2 / (2t)$ we obtain~\cite{supplemental} states with $f_0=1$, $p_0=\epsilon_0/(2t)$,
$p_{n \geq 1} = p_0 (2\xbar{E})^n/(\Pi_{m=1}^n \epsilon_m)^2$, $f_{n \geq 1} = 2p_{n-1}/\epsilon_n$. The 
dimers are destroyed, leaving single-peaked resonant states. While the localization length of these 
states follows the $t = 0$ case of Eq.\eqref{loclengthcorrdist=0} (Fig.\ref{fig:LL} red curves), the 
density of states behaves quite differently. 
First we note that the obtained states have positive $\xbar{E}$, 
which means that they must occur on the larger energy side of the FB energy. Furthermore, 
the density of states $\rho(E)$ follows~\cite{supplemental} from the PDF $f(z) = 1/(W\sqrt{z} )$ of the random number $z=\epsilon_0^2$ as
\begin{equation}
\rho(\xbar{E}) = \frac{1}{W} \sqrt{\frac{2t}{\xbar{E}}}  \;.
\label{doscorrdistneq0}
\end{equation}
The divergence is now strengthened to a square root one, but only on the high energy side of the FB energy. In Fig.~\ref{fig:DoS}, we indeed confirm this
singularity numerically on the right hand side of the FB energy. Meanwhile on the left hand side, we instead
observe a vanishing density of states. It should be also noted that we observe a gap developing as 
$t$ increases beyond a critical $t_c$ \cite{inprep}. This issue warrants further investigation, and may 
be related to disorder-induced crossing resonances \cite{ignatchenko}.

{\sl Mobility edges ---} Since the localization length is forced to vanish at the FB energy by correlated disorder in a one-dimensional system, it can be expected that a system with a metal-insulator transition will even have a singularity in
the mobility edge, i.e. the dependence of the critical potential strength on the eigenstate energy. Mobility edges typically appear for three-dimensional disordered systems, however a quasiperiodic potential is known
to produce a metal-insulator transition already in one space dimension. Indeed, a tight-binding chain
with eigenvalue problem $E\phi_n = \lambda \cos (2\pi \alpha n + \beta)\phi_n - (\phi_{n+1} + \phi_{n-1})$
is the well-known Aubry-Andr\'{e} model which has a metal-insulator transition at $\lambda_c=2$, provided
$\alpha$ is an irrational number~\cite{AA}. Note that $\lambda_c$ does not depend on the eigenenergy,
therefore the mobility edge function is a constant in the Aubry-Andr\'{e} case. In general, 
deviations from the Aubry-Andr\'{e} quasiperiodic case into other quasiperiodic potentials will lead
to the appearance of mobility edges \cite{soukoulis82, hiramoto89, zhou95, ananthakrishna97} - 
however, here we engineer them via a predictable analytical expression.

We again consider a correlated, quasiperiodic potential $\epsilon_n^a=-\epsilon_n^b=
\lambda \cos (2\pi \alpha n)$. Eq.\eqref{eq:tb} can be rewritten as
\begin{equation}
\widetilde{E} \; p_n = \widetilde{\lambda} \cos(4\pi\alpha n) \; p_n - (p_{n-1} + p_{n+1}),
\label{qpcorr}
\end{equation}
where
\begin{equation}
\widetilde{\lambda} = \dfrac{\lambda^2}{4(E-t)}, \quad \widetilde{E} := \dfrac{E + t}{2} - \dfrac{\lambda^2}{4(E-t)}.
\label{qpcorr2}
\end{equation}
%%
%%%%%%%%%%%%%%%%%%%%%%%%%%%%%%%%%%%%%%%%%%%%%%%%%%%%%%%%%%%%%%%
\begin{figure}[hbt]
 \centering
 \includegraphics[width=0.8\columnwidth]{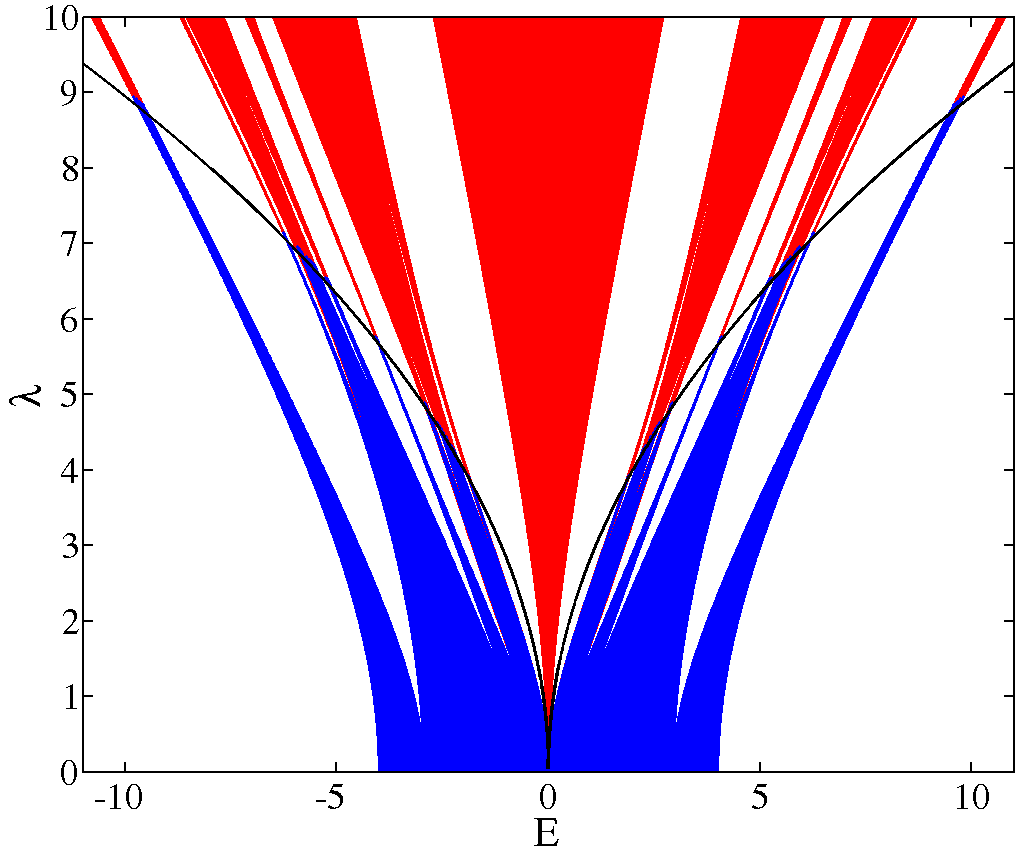}
 \caption{Spectrum of an $N=512$ unit cell chain under antisymmetric quasiperiodic perturbation with 
strength $\lambda$. The analytically predicted mobility edge Eq.\eqref{eq:ME} (black line) separates 
extended (blue) and localized (red; $\xi < 51$) modes.}
\label{fig:QP_Spectrum}
\end{figure}
%%%%%%%%%%%%%%%%%%%%%%%%%%%%%%%%%%%%%%%%%%%%%%%%%%%%%%%%%%%%%%%

Eq.\eqref{qpcorr} takes the form of a regular Aubry-Andr\'{e} model, however with effective energy 
$\widetilde{E}$ and potential strength $\widetilde{\lambda}$, which are functions of the eigenstate 
energy $E$ and the original potential strength $\lambda$, i.e. Eq.\eqref{qpcorr2}. Therefore if present, 
a metal-insulator transition must occur for $\widetilde{\lambda}=2$. This immediately yields a mobility 
edge dependence $\lambda_c(E)$:
\begin{equation}
\abs{\dfrac{\lambda_c^2}{4(E-t)}} = 2 \; \Rightarrow \; \lambda_c (E)= 2\sqrt{2|E-t|}\;.
\label{eq:ME}
\end{equation}
For $E=t$, the mobility edge curve is singular and zero, corresponding to the lack of any states, as 
previously mentioned. In Fig.~\ref{fig:QP_Spectrum} we show the spectrum of Eq.\eqref{qpcorr} as a 
function of $\lambda$.
We again compute the localization length $\xi(E, \lambda)$ with Eq.\eqref{eq:loclength}. If the recursion 
converges to a finite number (localized states, insulator), we plot blue points, while diverging cases 
are plotted in red (extended states, metal). The theoretical prediction Eq.\eqref{eq:ME} is also plotted 
and shows excellent agreement with numerical data.

{\sl Generalizations ---} Remarkably, this construction works in a plethora of other flat band models with CLS.
%, as long as the flat band can be decomposed into a (possibly overcomplete) set of compact localized modes. 
In higher dimensional lattices the construction of low energy eigenstates can proceed in exactly the same way: because the localization length is forced to vanish, for sufficiently small $\xbar{E}$ the eigenstates are near-sighted, so their properties are insensitive to the lattice dimension. The divergence in the density of states persists, in contrast to the more familiar van Hove singularities which get weaker as the dimension increases.  

\begin{figure}[hbt]
 \centering
 \includegraphics[width=0.8\columnwidth]{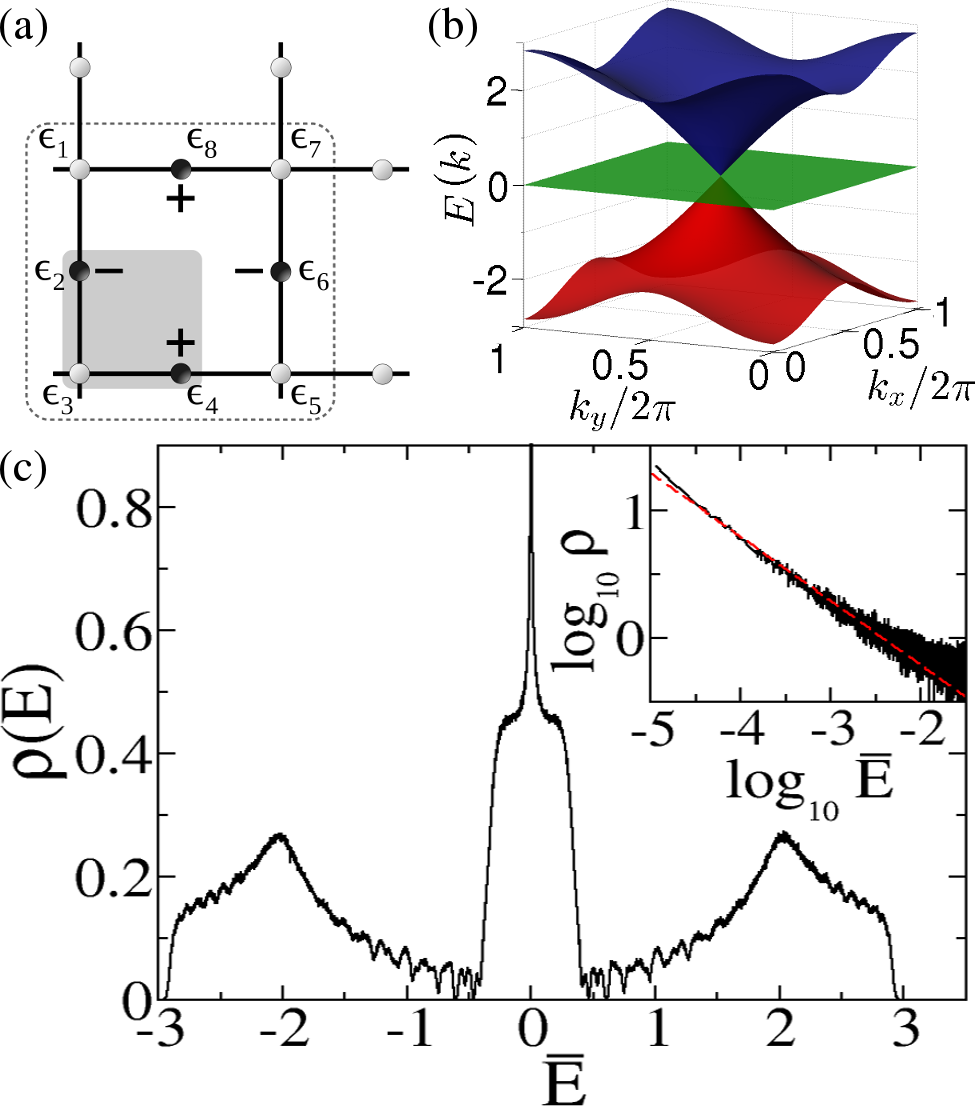}
 \caption{(a) The 2D Lieb lattice: its unit cell (shaded region), the 8-site plaquette (dashed enclosure),and the minimal compact state (black circles). 
(b) The band structure $E(k_x,k_y)$ from Eq.\eqref{2dlieb}. Red $E_-$ (bottom) and blue $E_+$ (top) bands 
are dispersive, the central $E_{FB}$ (green) band is flat.
(c) Density of states under the correlation $\epsilon_{2j} = (-1)^j \epsilon_a$ enforced at each plaquette, displaying square root singularity at $E = 0$. $W=1$. Lattice size is $N=24\times 24$ unit cells~\cite{lattice_size}. The red line is a linear fit.}
\label{fig:Lieb_Spectrum}
\end{figure}

As an example, we consider the 2D Lieb lattice, which hosts a flat band with nontrivial topology.
Here the compact localized states occupy multiple unit cells (shaded in Fig.~\ref{fig:Lieb_Spectrum}) and form an overcomplete non-orthogonal basis. 
Furthermore, the flat band is frustrated: its projector is long-ranged (power law decay in real space) and it is forced to touch another dispersive band~\cite{bergman2008, chalker10,lopes2011}.
The band structure is determined by two dispersive $E_{\pm}$ and one flat $E_{FB}$ bands~\cite{nita2013} (here all hoppings are
assumed to be of value unity):
\begin{equation}
E_{\pm}(k_x,k_y)= \pm 2\sqrt{ \cos^2 \frac{k_x}{2} + \cos^2 \frac{k_y}{2} }\;,\; E_{FB}=0\;.
\label{2dlieb}
\end{equation}
For a given CLS any onsite potential can be represented as a sum of a CLS-preserving part and its orthogonal counterpart.  A correlated potential for that given CLS is then defined by
zeroing the CLS-preserving part. Due to the above mentioned nontrivial topology of the 2D Lieb lattice, this procedure can be extended to every second CLS in a checkerboard
arrangement with unit cell coordinates $l_x=m+n$ and $l_y=m-n$ ($m,n$ are integers).
We realize the correlated potential by choosing $\epsilon_{2j} = (-1)^j \delta$ in each 8-site plaquette of a participating CLS (dashed enclosure) in Fig.~\ref{fig:Lieb_Spectrum}(a)
($\delta$ and the onsite energies $\epsilon_{2j-1}$ in the plaquette are random uncorrelated numbers with PDF $\mathcal{P}$).
Similar to the cross-stitch example, there are rapidly decaying eigenmodes with $E \sim \delta^2 \ll W^2 / 2$, which yield a square root singularity in the density of states. Fig.~\ref{fig:Lieb_Spectrum}(c) shows the corresponding numerical results~\cite{nogap}. The predicted square root singularity at $E=0$ lies on top of a background of width $W$ formed by the remaining CLS that have their energies renormalized. Also visible are two peaks at $E = \pm 2$, which are the van Hove singularities that have been regularized by the disorder. We note that the Lieb lattice was very recently fabricated as a photonic lattice using femtosecond laser writing~\cite{moti_lieb,diebel_lieb}. The required correlations can be readily introduced by modulation of the waveguide depths.

%%%%%%%%%%%%%%%%%%%%%%%%%%%%%%%%%%%%%%%%%%%%%%%%%%%%%%%%%%%%%%%
%%%%%%%%%%%%%%%%%%%%%%%%%%%%%%%%%%%%%%%%%%%%%%%%%%%%%%%%%%%%%%%

%Let us now generalize this construction. The cross-stitch flat band is ``topologically trivial'' in the sense that its Fano states are completely decoupled from the dispersive band via a local rotation. More interesting are ``nontrivial'' flat bands in which the Fano and dispersive states are entangled such that system cannot be adiabatically deformed to zero coupling limit and a local rotation cannot separate Fano and dispersive states. In this case, the compact Fano states occupy $U>1$ unit cells and form an overcomplete basis. But, one can still introduce a correlated potential that does not renormalize the energy of $100/U \%$ of the Fano states. By the same arguments, this will generate a singularity in the density of states (while remaining CLS will have their energy renormalized in some way).

%In higher dimensional lattices, one can construct low energy eigenstates in exactly the same way. Because $\xi \rightarrow 0$ as $\xbar{E} \rightarrow 0$ for any disorder, for sufficiently small $\xbar{E}$ the perturbative eigenstates are decaying so rapidly that the dimension of the lattice is irrelevant, so the singularity in the density of states will persist. In contrast, regular van Hove singularities become weaker as the lattice dimension $d$ is increased. 

%To confirm these arguments, we performed numerical calculations in $d=2$ lattices: Lieb, 2D cross-stitch. Indeed we observe singularities in density of states when appropriate potential is introduced, + mobility edges for quasiperiodic perturbation.

{\sl Conclusion ---} We have shown how appropriately correlated disorder can transform the singular 
density of states at a flat band into weaker logarithmic or square root divergences. The resulting 
simple, analytically tractable models feature vanishing localization lengths for arbitrarily weak 
disorder, and mobility edges for quasiperiodic perturbations. This approach offers a flexible and 
intuitive way to engineer different types of spectral singularities or mobility edges in lattice systems 
and control wave transport.

%In summary, we have shown how to generate singularities in the density of states, vanishing localization length for arbitrarily weak disorder, mobility edges, etc. even in simple one-dimensional models by applying locally correlated perturbations to flat bands. The local correlations preserve the energies of compact localized flat band modes while enforcing their hybridization with other dispersive states. Competition between the macroscopic number of flat band states is responsible for this unusual behavior, which readily generalizes to higher dimensions. This procedure offers a flexible and intuitive way to engineer different types of spectral singularities in lattice systems...

\newpage
\setcounter{equation}{0}
\renewcommand{\theequation}{S\arabic{equation}}
%%%%%%%%%%%%%%%%%%%%%%%%%%%%%%%%%%%%%%%%%%%%%%%%%%%%%%%%%%%%%%%%%%%%%%%%%%%%%
\section{Supplemental Material}
This Supplemental Material presents derivations of the localization length Eq.(6), the density of states 
Eqs.(7,8), and the profiles of the low energy eigenstates appearing in the main text.

{\sl Localization Length ---}
For $\epsilon_n^+ = 0$, $f_n$ can be eliminated from the eigenmode equations Eq.(3), leaving
\begin{equation}
\left( \frac{(\epsilon_n^-)^2}{\xbar{E}} - \xbar{E} - 2 t  \right) p_n = 2 (p_{n-1} + p_{n+1} ).
\end{equation}
When $\xbar{E} \ll W^2 / 4$ is small, the first term on the left hand side is resonantly enhanced and 
dominates. The ratio $R_n = p_{n+1} / p_n$ is approximated by
\begin{equation}
R_n \approx \frac{(\epsilon_n^-)^2}{2 \xbar{E}} - \frac{1}{R_{n-1}},
\end{equation}
The decaying solution for small $\xbar{E}$ is $R_{n-1}(\epsilon_n^-) \approx 2 \xbar{E} / 
(\epsilon_n^-)^2$, thus applying Eq.(5) we obtain
\begin{align}
\xi^{-1} &= \lim_{M \rightarrow \infty} \frac{1}{M} \sum_{n=1}^M \ln \left|\frac{2 
\xbar{E}}{(\epsilon_n^-)^2}\right|, \nonumber \\
&= \langle \ln \left| \frac{2 \xbar{E}}{(\epsilon_n^-)^2} \right| \rangle.
\end{align}
$\epsilon_n^-$ are uncorrelated random variables with a uniform  probability distribution function (PDF),
\begin{equation} 
f_{\epsilon} ( x ) = \begin{cases} \frac{1}{W},& \text{if } |x| \leq \frac{W}{2} \\
0, & \text{otherwise} \end{cases}
\end{equation}
thus the disorder average is
\begin{align}
\xi^{-1} &= \frac{1}{W} \int_{-W/2}^{W/2} \ln \left|\frac{2 \xbar{E}}{x^2} \right| dx \nonumber \\
&= 2 + \ln \left| \frac{8 \xbar{E}}{W^2} \right|,
\end{align}
which reproduces Eq.(6) (noting that $\xbar{E}/W^2 \ll 1$ and taking $\xi$ to be positive).

Curiously, Eq.(6) incorrectly predicts $\xi^{-1} = 0$ at $\xbar{E}/W^2 = 1/(8e^2) \approx 0.02$, well 
within the validity of the approximation $\xbar{E}/W^2 \ll 1/4$. To explain this anomaly, we note that 
the perturbative result Eq.(S2) is only valid when $E/(\epsilon_n^-)^2 \ll 1 \Rightarrow \epsilon_n^- 
\gg \sqrt{E}$. Thus, the integral in Eq.(S5) requires a finite cutoff $a \sim \sqrt{E}$
\begin{equation}
\xi^{-1} = \frac{2}{W} \left( \int_0^a \ln \abs{R(x)} dx + \int_a^{W/2} \ln 
\abs{\frac{2\xbar{E}}{x^2}} dx \right), 
\end{equation}
and we require $a \ll W/2$ for the first term to be negligible. Thus, Eq.(6) is only a good approximation
under the stricter condition $\sqrt{E}/W \ll 1$, which excludes the divergence of the localization 
length, $\xi^{-1} = 0$, at $\sqrt{E}/W \approx 0.13$. 

{\sl Density of States ---}
To obtain the density of states Eq.(7), we evaluate the PDF of the random variable $z = \epsilon_0^- 
\epsilon_1^-$. 
The product distribution $f_{z}(x)$ is given by
\begin{align}
f_z (x ) &= \int f_{\epsilon} (y) f_{\epsilon}(x/y) \frac{1}{|y|} dy, \nonumber \\
&= \frac{1}{W} \int_{-W/2}^{W/2} \frac{1}{|y|} f_{\epsilon} (x / y) dy, \nonumber \\
&= \frac{2}{W^2} \int_{2|x|/W}^{W/2} \frac{dy}{y}, \nonumber \\
&= \begin{cases} \frac{2}{W^2} \left[ \ln \frac{W}{2} - \ln \frac{2 |x|}{W} \right], & \text{if } |x| 
\le 
\frac{W^2}{4}, \\
0, & \text{otherwise} \end{cases}
\end{align}
Eq.(7) follows by making the change of variables $E = z/2$, with $\rho ( E ) = 2 f_z (2 E)$. 

Similarly, we obtain Eq.(8) from the PDF of $z = \epsilon_0^2 = g ( \epsilon_0)$ via
\begin{equation}
f_z ( x ) = 2 |\partial_x g^{-1} ( x ) | f_{\epsilon} ( g^{-1} ( x ) ),
\end{equation}
where $g^{-1} ( x ) = \sqrt{x}$. This yields
\begin{equation}
f_z ( x ) = \begin{cases} \frac{1}{W\sqrt{x}},& \text{if } 0 < x < \frac{W^2}{4}, \\ 0, & 
\text{otherwise} \end{cases}
\end{equation}
which gives Eq.(8) after the change of variables $E = z / (2t)$.

By the same arguments as above, the incorrectly predicted vanishing of $\rho(\xbar{E})$ at 
$\xbar{E}=W^2/4$ occurs due to realizations of the potential outside the range of validity
of the perturbative expansion, and instead the stricter condition $\sqrt{E}/W \ll 1$ is again
required.

{\sl Low Energy Eigenstates ---}
The initial conditions $f_{0,1}$ uniquely determine the eigenmode amplitude along the rest of the 
lattice. The eigenmode equations for sites $p_{0,1}$ read
\begin{align}
\left(\frac{\epsilon_0^- }{2} -\frac{\xbar{E}^2}{2\epsilon_0^-} - \frac{t \xbar{E}}{\epsilon_0^-} 
\right) 
f_0 &= p_{-1} + \frac{\xbar{E} f_1}{\epsilon_1^-}, \\
\left( \frac{\epsilon_1^- }{2} -\frac{\xbar{E}^2}{2\epsilon_1^-} - \frac{t \xbar{E}}{\epsilon_1^-} 
\right) f_1 &= p_{2} + \frac{\xbar{E} f_0}{\epsilon_0^-}.
\end{align}
Without loss of generality, we can set $f_0 = 1$. When $\xbar{E}$ is small, from the calculation of the 
localization length we have $p_{-1,2} \approx 2\xbar{E} \, p_{0,1} / (\epsilon_{-1,2}^- )^2 \approx 0$. 
Under this approximation, the above equations are solved to leading order in $\xbar{E}$ to obtain, for 
$t 
= 0$,
\begin{equation}
\xbar{E} = \pm \epsilon_0^- \epsilon_1^- / 2, f_1 = \pm 1,
\end{equation}
and when $t \ne 0$
\begin{equation}
\xbar{E} = \epsilon_0^2 / (2t), f_1 = \epsilon_0^{-} / (t \epsilon_1^{-} ),
\end{equation}
which yield the eigenmode profiles appearing in the main text. To verify this result we also obtained 
eigenstates numerically for various realizations of disorder. The small $\xbar{E}$ eigenstates indeed 
display a single strong maximum, with energy determined by disorder potential at this maximum according 
to the above equations.


\begin{thebibliography}{}
\bibitem{inprep} The value observed is $t_c > 1$, however in such a gap we see no indication 
of Lifshitz-Urbach tails signifying rare states. Since our understanding is based solely on numerics, we 
can not completely exclude these rare events, and therefore can not provide an exact value to $t_c$. 
Further investigation into gap development is ongoing.
\bibitem{nogap}
There is no gap in this case: disorder breaks the particle-hole symmetry, but the ensemble average restores it.
\bibitem{lattice_size}
Because $\xi \rightarrow 0$ at $E_{FB}$, this is sufficient to avoid finite size effects. We checked that increasing the lattice size does not significantly change the results.
\end{thebibliography}
\end{document}